\documentclass[prb,aps,epsfig,twocolumn,showpacs]{revtex4-1}
\usepackage{color}
\usepackage{epsfig,amssymb,amsmath,latexsym}
\usepackage{float}
\usepackage{color}
\usepackage[colorlinks=true,linktoc=page,linkcolor=blue,citecolor=magenta]{hyperref}

\newcommand\bea{\begin{eqnarray}}
\newcommand\eea{\end{eqnarray}}
\newcommand\beq{\begin{equation}}  
\newcommand\eeq{\end{equation}}

\newcommand{\non}{\nonumber}  

\newcommand\ie{{\it{i.e.}}}

\begin{document}
\textheight=23.8cm

\title{Tunneling Conductance in Normal-Insulator-Superconductor junctions of Silicene}
\author{Surajit Sarkar$^{1}$, Arijit Saha$^{2,3}$ and Suhas Gangadharaiah$^{1}$}
\affiliation{\mbox{$^1$} {Department of Physics, Indian Institute of Science Education and Research, Bhopal, India} \\
\mbox{$^2$}{Institute of Physics, Sachivalaya Marg, Bhubaneswar, Orissa, 751005, India}\\
\mbox{$^3$}{Homi Bhabha National Institute, Training School Complex, Anushakti Nagar, Mumbai 400085, India}
}

\date{\today}

\begin{abstract}
We theoretically investigate the transport properties of a normal-insulator-superconductor (NIS) junction of silicene in
 the thin barrier limit.  Similar to graphene  the tunneling conductance in such NIS structure exhibits an oscillatory 
behavior as a function of the strength of the barrier in the insulating region. However, unlike in graphene, the tunneling conductance in silicene can be 
controlled by an external electric field owing to  its  buckled structure. We also demonstrate the change in behavior of 
the tunneling conductance across the NIS junction as we change the chemical potential  in the normal silicene region. 
In addition, at high doping levels in the normal region, the period of oscillation of the tunneling conductance as a function of the 
barrier strength changes from $\pi/2$ to $\pi$ with the variation of doping in the  superconducting region of silicene. 
\end{abstract}

\maketitle

\section{Introduction}

One of the most active research field in condensed matter physics  since the  last decade   has been the
study of Dirac fermions in graphene~\cite{geimreview} and topological insulator~\cite{sczhangreview,hasan2010colloquium}. 
The low energy spectrum of these materials satisfies massless Dirac equation.   The relativistic band structure of the Dirac fermions has lead to  tremendous  interest in graphene in  terms of possible application as well as from the point of view of fundamental physics.

Very recently, a silicon analogue of graphene, silicene has been attracting a lot of attention both theoretically 
and experimentally~\cite{MEzawaReview,CCLiu1,siliceneexp1,siliceneexp2,siliceneexp3}, due to the possibility of new applications, 
given its compatibility with silicon based electronics. 
Unlike graphene,  silicene does not have a planar structure;  instead the
buckled structure of silicene manifests itself  as a  spin-orbit coupling resulting in a band gap at the Dirac point~\cite{CCLiu1}. 
More interestingly,  it has been reported earlier that such band gap is tunable by an external electric field applied perpendicular to the silicene 
sheet~\cite{Drummond,MEzawa3}. This opens up the possibility of realizing silicene based electronics and very recently a silicene
based transistor has been experimentally realized~\cite{LTao}.

In recent times, it has been realized that topologically non-trivial phases arise in silicene, tuned by the external 
electric field only~\cite{MEzawa3,MEzawa1,MEzawa5}. Graphene and silicene have similar band structures and
the low energy spectrum of both are described by the relativistic Dirac equation \ie, both have the Dirac cone
band structure around the two valleys represented by the momenta {$ \textbf{K}$} and $\textbf{ K}^{\prime}$.
However, the important difference between graphene and silicene is that the spin-orbit coupling (SOC) in silicene 
is much stronger than in graphene~\cite{CCLiu1,MEzawa3,GianG} which causes the Dirac fermions in silicene
to become massive. Furthermore, due to the buckled structure in silicene, the two sub-lattices respond differently
to an externally applied electric field resulting in electrically tunable Dirac mass term~\cite{MEzawa3}. 
Such tunability allows for the mass gap to be closed at some critical value of the electric field and then reopened.
Hence, the phases on the two sides of the critical electric field where the gap is closed are different, with one of 
them being topologically trivial and the other being topologically non-trivial~\cite{MEzawa3,MEzawa1,MEzawa5}. 
As a result, silicene under the right circumstances can be a quantum spin hall insulator with topologically 
protected edge states~\cite{CCLiu2,MEzawa1,MEzawa2}.

The advent of superconductivity in  graphene and certain topological insulators via the proximity effect 
has led to an upsurge of interest in this area~\cite{beenakkerreview,sczhangreview}.
Very recently, superconducting proximity effect
in silicene has been reported in Ref.~\onlinecite{LinderYokoyama} where the authors have investigated the behavior of 
Andreev reflection (AR) and crossed Andreev reflection (CAR) in a normal-superconductor (NS) and normal-superconductor-normal (NSN)
junctions of silicene respectively. 

In this article, we study the behavior of tunneling conductance (TC) in a normal-insulator-superconductor (NIS) junction 
of silicene where superconductivity in silicene is induced via the proximity effect. We model our NIS setup within the 
scattering matrix formalism~\cite{beenakker1,blonder1982transition} and obtain the external electric field controllable
TC for  thin  barrier limit. Similar set up in graphene have been studied earlier in Refs.~\onlinecite{subhro,moitri1}. 
However, TC based on silicene NIS structure has not yet been considered in the literature.

The remainder of this paper is organized as follows. In Sec.~\ref{sec:II}, we present our model for the silicene NIS structure
and describe the scattering matrix formalism to compute the tunneling conductance. In Sec.~\ref{sec:III},
we present our results for the TC in the NIS set-up for the thin barrier case. Finally in Sec.~\ref{sec:IV},
we summarize our results followed by the conclusions. 

\section{Model and Method} {\label{sec:II}}
\begin{figure}
\centering
\includegraphics[width=1.0\linewidth]{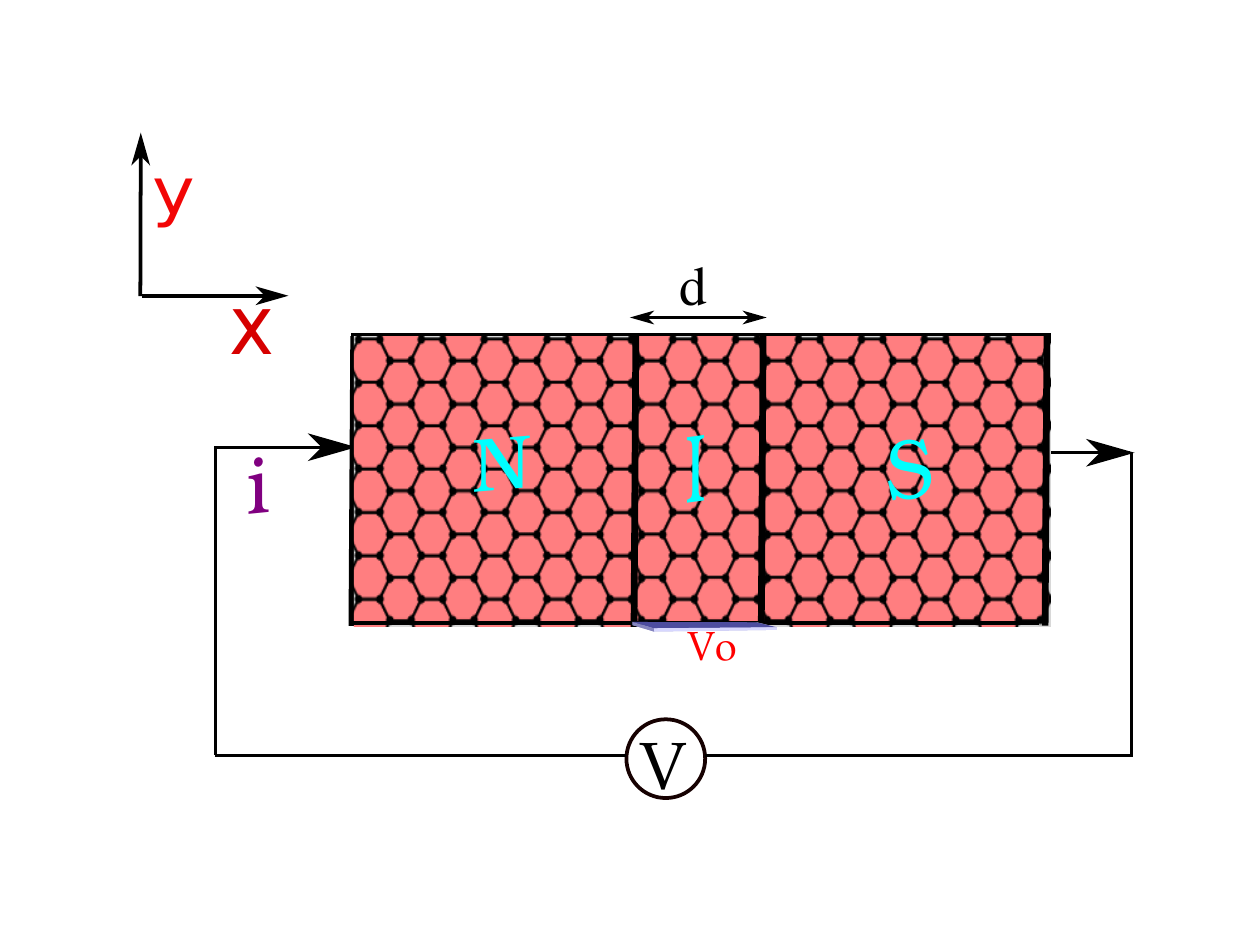}
\caption{(Color online) Schematic of the silicene NIS junction on an $xy-$plane.  $N$ corresponds to the  normal region, a barrier of height $V_0$ is applied in the $I$ region (with width $d$) to make it insulating. Superconductivity is induced in the $S$ region  via the proximity effect.   }
\label{setup}
\end{figure}


In this section we will set up the equations  to study the  transport properties of  an NIS  junction in a silicene sheet placed along the $xy$-plane  (see Fig.~\ref{setup}).   
The  $x\leq -d$ region is the normal  region (N), the  insulating region (I) has a  width $d$ and occupies the  $-d\leq x\leq 0$ region, while the  superconducting region (S) 
occupies $x\geq 0$ region. The insulating region has a gate tunable barrier potential of strength $V_0$, while the
superconductivity in  $x\geq 0$ region  is assumed to have been  induced via the proximity effect  where the external superconductor is taken to be of the  
 s-wave type. 

 The  Silicene NIS junction is  described by the Dirac Bogoliubov-de Gennes (DBdG) equation of the form
\begin{equation}{\label{BdG-Hamil}}
\begin{pmatrix}
{\hat{H_\eta}}&{\Delta}{\hat{I}}\\
{\Delta}^{\dagger}{\hat{I}} & -{\hat{H_\eta}}\end{pmatrix}{\Psi}=E{\Psi},
\end{equation}
where  $E$ is the excitation energy, $\Delta$ is the proximity induced superconducting energy gap.  
The Hamiltonian $\hat{H}_\eta$ describes the low energy physics near each of the $\textbf{K}$, $\textbf{K}^\prime$ Dirac points  and has the form~\cite{LinderYokoyama},
\begin{equation}
H_\eta=\hbar v_F(\eta k_x \hat{\tau}{_x}-k_y \hat{\tau}{_y})+(lE_Z-\eta\hat{\sigma}_z\lambda{_S}{_O})\hat{\tau}{_z}-\mu{_i}\hat{1},
\label{hamilt}
\end{equation}
where $\eta=+(-)$ corresponds to $\textbf{K}$ ($\textbf{K}^\prime$) valleys,  $v_F$ is the fermi velocity (in the following we will set $\hbar v_F=1$),  $\mu{_i}$ ($i$  represents any of  the  N/I/S regions) is the chemical 
potential and $\lambda_{SO}$ is the spin orbit term. The Pauli matrices $\hat{\sigma}$ and $\hat{\tau}$ act on the spin and sub-lattice space, respectively.
The  potential energy term  $l E_Z$ owes its origin to the buckled structure of the Silicene  wherein  
the $A$ and $B$ sites occupy  slightly different planes (separated by a distance of length  $l$) and therefore 
acquire a potential difference when an external electric field $E_Z$
is applied perpendicular to the plane.  It turns out that at the critical electric field  $E^c_Z=\lambda_{SO}/l$
each of the valleys become gapless  with   the gapless bands of one of  the valley  being  up-spin polarized and the other
  down-spin polarised\cite{MEzawa3,ruchi}. Away from the critical field, the bands  (corresponding to $H_\eta$) at each of the  $\textbf{K}$ and $\textbf{K}^\prime$  points are split into
  two conduction and two valence bands with the band gap being  $|lE{_Z}-{\eta}{\sigma}{\lambda}{_S}{_O}|$, where $\sigma =\pm 1$ is a spin index.

Assuming translational invariance along the $y$-direction we solve
Eq.~(\ref{BdG-Hamil}) to find the wave functions in all the three different regions.
The wave functions for the electrons and holes moving along the $\pm x$ direction in the N region are 
 \begin{eqnarray}
\psi_N^e{^\pm}&=&\sqrt{\frac{1}{2\tau^e_1(E+{\mu_N})}}
          \begin{bmatrix}
         k^e_1e^{\pm}{^i}{^\eta}{^{\alpha_e}} \\ \pm \eta\tau^e_1 \\0 \\ 0
         \end{bmatrix}e^{i({\pm}k^e_{1x}x+k^e_{1y}y)}\notag\\
\psi_N^h{^\pm}&=&\sqrt{\frac{1}{2\tau^h_1(E-{\mu_N})}}
          \begin{bmatrix}
      0 \\ 0 \\{k^h_1e^{\pm}{^i}{^\eta}{^{\alpha_h}}}\\ \mp \eta\tau^h_1
         \end{bmatrix}e^{i({\pm}k^h_{1x}x+k^h_{1y}y)},
\end{eqnarray}
 where  $\,{\tau^{e(h)}_1}=E{\pm}{\mu_N}{\mp}(lE{_Z}-{\eta}{\sigma}{\lambda}{_S}{_O})$ and $E$ is the energy of the particle wrt. to the Fermi level $\mu_N$.  
  We  note that due to the spin being a good quantum number (and also because of time reversal symmetry) we can restrict our discussion by considering spin of only one type.

The conservation of momentum  along the $y$-direction, $k_{1y}^{e} = k_{1y}^{h} $, leads to the  angle of incidence ${\alpha_e}$  and the Andreev reflection   angle ${\alpha_h}$ being related  via, $k^h_1\sin({\alpha{_h}})=k^e_1\sin({\alpha{_e}})$  where
\beq
 k^{e(h)}_1={\sqrt{(E{\pm}{\mu{_N}})^2-(lE{_Z}-{\eta}{\sigma}{\lambda}{_S}{_O})^2}}.
\eeq
 In the insulating region the wave functions are 
 \begin{eqnarray}
 \psi_I^e{^\pm}&=&\sqrt{\frac{1}{2\tau^e_2(E+{\mu_I})}}
           \begin{bmatrix}
          k^e_2e^{\pm}{^i}{^\eta}{^{\beta_e}} \\ \pm \eta\tau^e_2 \\0 \\ 0
          \end{bmatrix}e^{i({\pm}k^e_{2x}x+k^e_{2y}y)}\notag\\
 \psi_I^h{^\pm}&=&\sqrt{\frac{1}{2\tau^h_2(E-{\mu_I})}}
            \begin{bmatrix}
       0 \\ 0 \\k^h_2e^{\pm}{^i}{^\eta}{^{\beta_h}}\\ \mp\eta\tau^h_2
          \end{bmatrix}e^{i({\pm}k^h_{2x}x+k^h_{2y}y)},\notag\\
 \end{eqnarray}
where  ${\tau^{e(h)}_2}=E{\pm}{\mu_I}{\mp}(lE{_Z}-{\eta}{\sigma}{\lambda}{_S}{_O})$ and
\begin{eqnarray}
k^{e(h)}_2={\sqrt{(E{\pm}{\mu{_I}})^2-(lE{_Z}-{\eta}{\sigma}{\lambda}{_S}{_O})^2}},
\end{eqnarray} 
 where $\mu_I=\mu_N-V_0$ and  $V_0$ is the electrostatic potential that controls the barrier height.

 Finally, in the superconducting region the wave functions of DBdG quasiparticles are given by,
\begin{eqnarray}
   \psi_S^e&=&\frac{1}{\sqrt{2}}
             \begin{bmatrix}
           u{_1} \\ {\eta}u{_1}e^i{^\eta}{^{\theta_e}} \\u{_2}\\{\eta}u{_2}e^i{^\eta}{^{\theta_e}}
            \end{bmatrix}e^{(i{\mu}{_S}-{\kappa})x+iq^e_yy}\notag\\
   \psi_S^h&=&\frac{1}{\sqrt{2}}
             \begin{bmatrix}
        u{_2}\\-{\eta}u{_2}e^-{^i}{^\eta}{^{\theta_h}}\\u{_1}\\-{\eta}u{_1}e^-{^i}{^\eta}{^{\theta_h}}
            \end{bmatrix}e^{(-i{\mu}{_S}-{\kappa})x+iq^h_yy},
\end{eqnarray}
  where 
  \begin{eqnarray}
  u_{1/2}= \sqrt{ \frac{1}{2} \pm \frac{\sqrt{E^2-\Delta^2}}{2E}}~~\text{and}~~ \kappa=\sqrt{\Delta^2-E^2}.
  \end{eqnarray}
 As before, momentum conservation along the $y$-direction relates  the transmission angles for electron-like and hole-like quasi-particles via the following 
equation
\beq
q^i\sin{\theta}_{i}=k^e_1\sin{\alpha}_{e} ,
\label{sc1}
\eeq
 for $i=e,h$. The quasiparticle momentums are given by
\beq
 q^{e(h)}={\mu}{_S}\,{\pm}\,{\sqrt{E{^2}-{\Delta}{^2}}}
\label{sc2}
\eeq
where $\mu_S=\mu_N+U_0$ and $U_0$ is the gate potential applied in the superconducting region to tune the Fermi surface mismatch.
 
 Let us  consider an electron with energy  $E$ incident on the interface of a conventional NIS junction
 of a silicene sheet.  Part of the wave function gets transmitted and the rest undergoes  both normal 
 and Andreev reflection from the interface. Taking into consideration all these processes the  wave
 functions in the different regions of junction can be written as~\cite{blonder1982transition}:
\begin{eqnarray}
\Psi_N&=&\psi_N^{e+}+r\psi_N^{e-}+r{_A}\psi_N^{h-}\non\\
\Psi_I&=&p\psi_I^e{^+}+q\psi_I^e{^-}+m\psi_I^h{^+}+n\psi_I^h{^-}\non\\
\Psi_S&=&t{_e}\psi_S^e+t{_h}\psi_S^h,
\label{bc}
\end{eqnarray}
where $r$ and $r{_A}$ are the amplitudes of normal reflection and Andreev reflection in the $N$ region, respectively. The  transmission amplitudes $t{_e}$ and $t{_h}$ correspond to  electron like and hole like quasiparticles in the $S$ region, respectively. From the continuity of the wave functions  at the two interfaces we have
\begin{eqnarray}
\Psi_N|{_{x=-d}}  =  \Psi_I|{_{x=-d}} \,\,\,\,\,\,\,\,\,\,\,\,\,\,\,\, \Psi_I|{_{x=0}}=\Psi_S|{_{x=0}} \label{Boundary-Cond}
\end{eqnarray}
which yields a set of eight linearly independent equations. Solving them we obtain $r$ and $r_A$,   these amplitudes fully determine  the 
tunnelling  conductance of the NIS junction within the 
Blonder-Tinkham-Klapwijk 
formalism (BTK) ~\cite{blonder1982transition}
\begin{eqnarray}
{\frac{G(eV)}{G{_0}(eV)}}=\int_{0}^{\pi/2}\Bigg[1-|r|{^2}+|r{_A}|{^2}{\frac{\cos({\alpha}{_h})}{\cos({\alpha}{_e})}}\Bigg]\cos({\alpha}{_e})d{\alpha}{_e}\nonumber\\
\end{eqnarray} 
where  $G{_0}$ is the ballistic conductance of silicene.

\section{Thin barrier}{\label{sec:III}}
In this section we will solve the scattering problem in the limiting  case of thin barrier. We will consider  the limit where the barrier height $V_0\rightarrow{\infty}$
while the width $d\rightarrow0$  such that the product $V_0 d\rightarrow \chi$ is finite and non-zero\cite{subhro}. 
We will consider only those scenarios wherein  the  mean-field criterion  for superconductivity, i.e., $\mu_S=\mu_N +U_0 \gg \Delta$, is satisfied.  This can be achieved by 
controlling either the doping level $\mu_N$ or the gate voltage $U_0$ in the superconducting region.
Solving the boundary condition (\ref{Boundary-Cond}) for 
 $r$ and $r_A$  we obtain,
\begin{eqnarray}
r=\frac{N{_R}}{D} ,\,\,\,\,\,\ r_A=\frac{N{_{AR}}}{D},
\end{eqnarray}
where
 \begin{eqnarray}
 D=-e^{4i\chi}(e^{i\alpha{_e}}-\gamma)(e^{i\alpha{_h}}-\delta)+e^{2i\beta}(e^{i\alpha{_e}}+\gamma)(e^{i\alpha{_h}}+\delta),\nonumber
 \end{eqnarray} 
 \begin{eqnarray}
 N{_R}&=&e^{i\alpha{_e}}\big[e^{4i\chi}(1+{\gamma}e^{i\alpha{_e}})(e^{i\alpha{_h}}-\delta)+e^{2i\beta}({\gamma}e^{i\alpha{_e}}-1)\nonumber\\
&&\times  (e^{i\alpha{_h}}+\delta)\big],\nonumber
 \end{eqnarray}
  and 
 \begin{eqnarray}
 N{_{AR}}=2{\tilde{A}}e^{i(\alpha{_h}+2{\chi})}(1+e^{2i\alpha{_e}}){\gamma}e^{i\beta}.
 \end{eqnarray}
 The remaining  parameters are defined as follows, 
 \begin{eqnarray}
{\gamma}={\sqrt{\frac{({\epsilon}+{\mu}{_N}+{\lambda})}{({\epsilon}+{\mu}{_N}-{\lambda})}}};\,\,\,
{\delta}={\sqrt{\frac{({\epsilon}-{\mu}{_N}-{\lambda})}{({\epsilon}-{\mu}{_N}+{\lambda}})}}\nonumber\\
{\tilde{A}}={\sqrt{\frac{({\epsilon}+{\mu}{_N}-{\lambda})({\epsilon}-{\mu}{_N})}{({\epsilon}-{\mu}{_N}+{\lambda})({\epsilon}+{\mu}{_N})}}}\,\,\,\,\,\,\,\,\,\,\,\,\,\,\,\,\,\,
\end{eqnarray}
 where      $e^{i\beta}=u{_1}/u{_2}$, $\epsilon=eV/\Delta$ and $\lambda \Delta=(lE{_Z}-{\eta}{\sigma}{\lambda}{_S}{_O})$.
 Unless   otherwise stated we consider a scenario wherein the product ${\eta}{\sigma}=1$, and the  upper conduction sub-band remains unfilled.

\begin{figure}[h]
\centering
\includegraphics[width=1.0\linewidth]{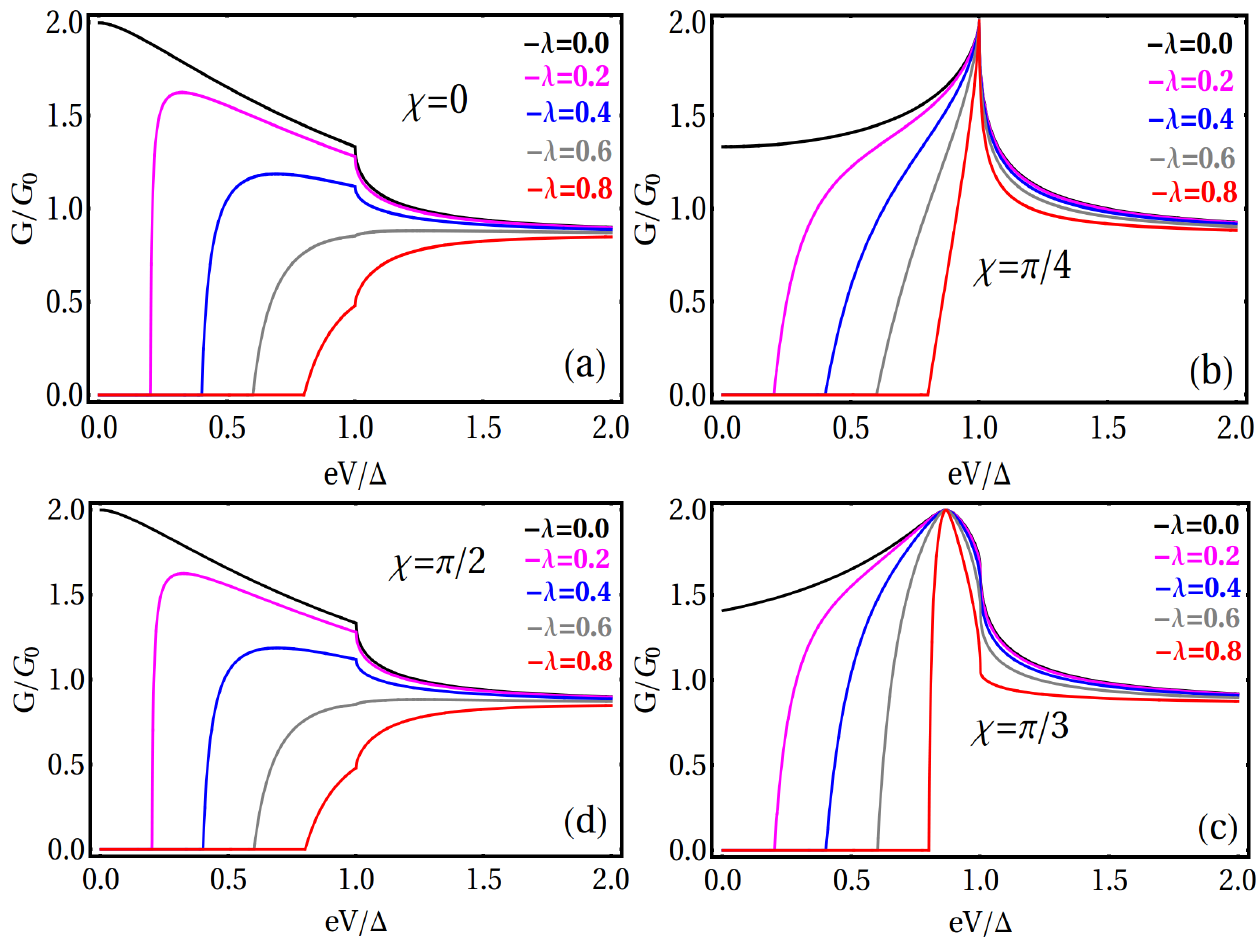}
\caption{(Color online)  Plot of tunneling conductance of a NIS silicene junction as a function of bias voltage  for different barrier strength $\chi$ and gap $\lambda$ in the  undoped regime ($\mu_{N}=0$) and for $U{_0}\gg \Delta$.}
\label{GvseV}
\end{figure}
\begin{figure}
\centering
\includegraphics[width=1.0\linewidth]{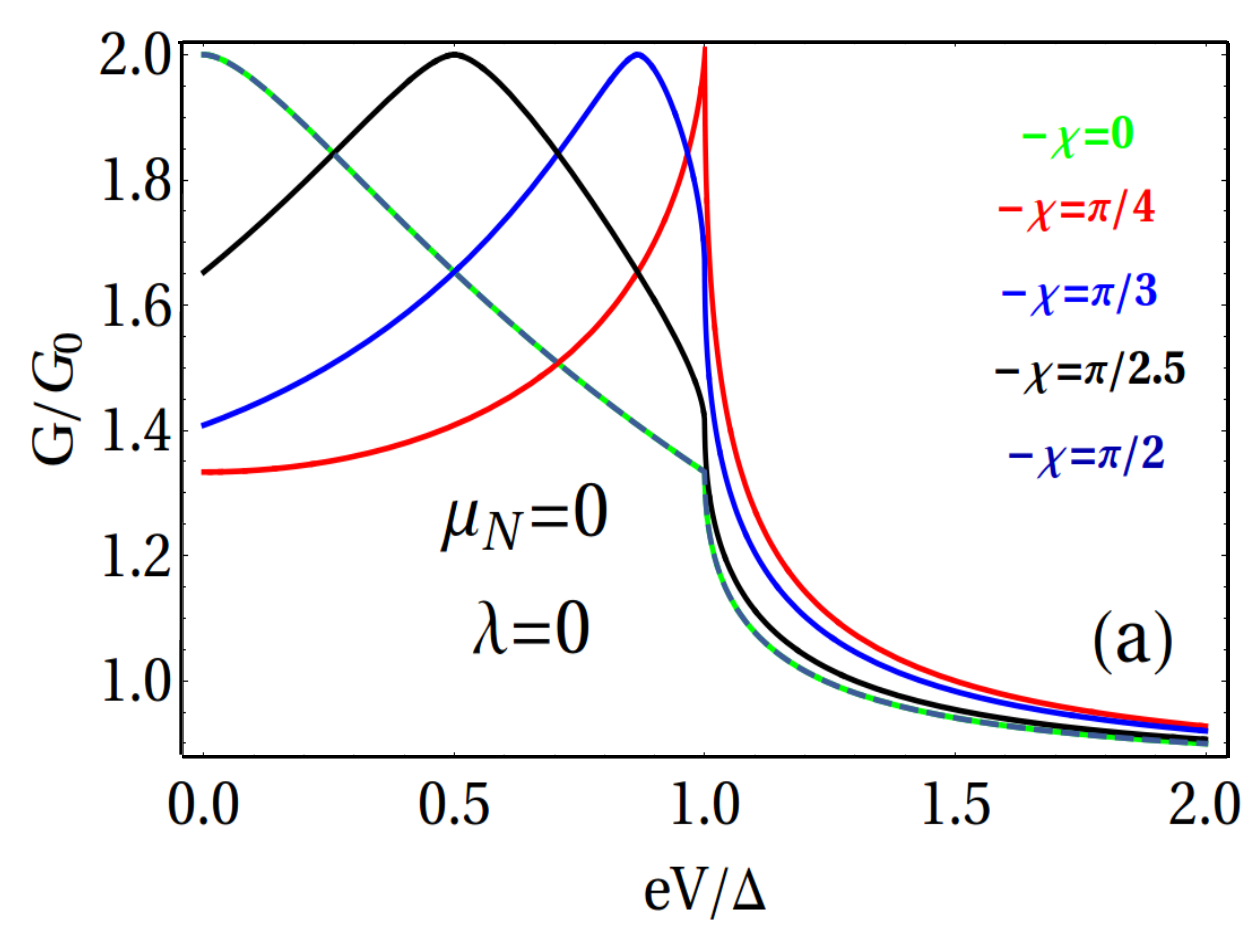}
\includegraphics[width=1.0\linewidth]{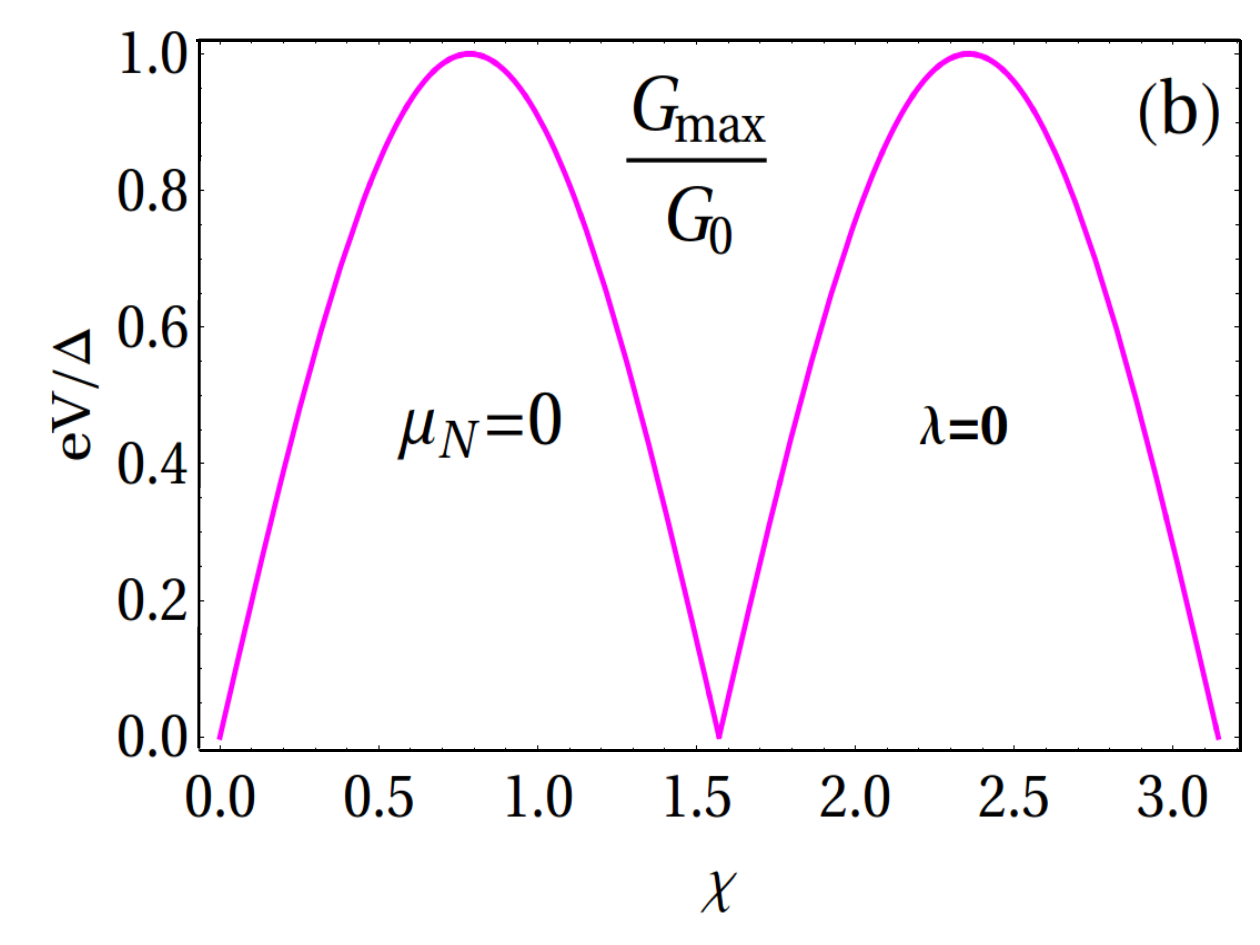}
\caption{(Color online) (a)Tunneling conductance as a function of bias voltage for different barrier strength at ${\lambda}=0$ and $U{_0}\gg \Delta$. (b) 
The position of  the maxima of $G/G{_0}$ in the $eV/\Delta$ vs $\chi$  plot.}
\label{Two_figures}
\end{figure}

\begin{figure}[h]
\centering
\includegraphics[width=1.0\linewidth]{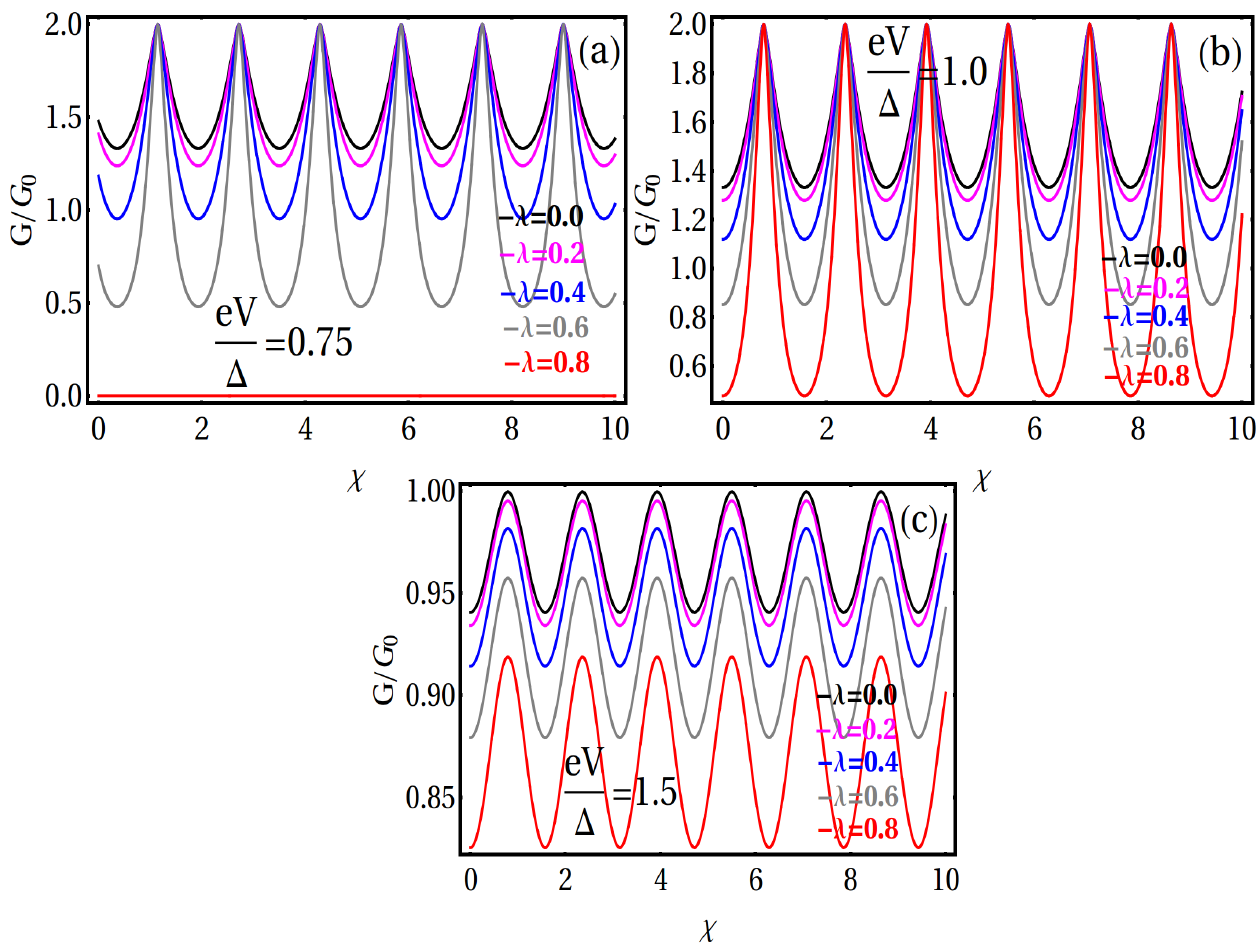}
\caption{(Color online) Tunneling conductance in the limit of $U{_0}\gg \Delta$ as a function of barrier strength ${\chi}$ for undoped regime ($\mu_{N}=0$),  (a) in the sub-gap regime, (b)  $eV=\Delta$ and, (c) $eV > \Delta$.}
\label{Gvschi}
\end{figure}
\subsection{Undoped regime}

In this subsection we  focus on the undoped regime ($\mu_{N}=0$)     in the normal side of the silicene sheet.   
The tunnel conductance value $G/G_0$ varies from $0$ to a peak value of $2$ with the Andreev reflection contributions being purely of the specular kind. 
The plot of  $G/G_0$ as a function of  $eV/\Delta$ for a fixed  barrier
 strength ${\chi}$  and different $\lambda$'s  are shown in Fig.~\ref{GvseV}.  For the    
transparent  barrier regime  (i.e., ${\chi}=0$) we obtain plots,  Fig.~\ref{GvseV}a,  identical to those in  
Ref.~(\onlinecite{LinderYokoyama}). 
A common theme for $G/G_0$ vs $eV/\Delta$ for all the different barrier strengths [see plots Figs.~\ref{GvseV}(a)-(d)]  is that the  
non-zero contribution to conductance arises when the incident electron has energy greater than the band gap.  
 
Interestingly, this feature can be exploited by the electric field applied perpendicular to the plane which can then be used as a 
switch to turn on or off the conductance.  Tuning the  
electric field so that $\lambda=0$ brings the system to the Dirac limit and assures non-zero conductance  at zero bias and for  
arbitrary strengths of the barrier. 
 Another similarity between the different plots  is the significant  change
 in the slope at $eV/\Delta =1$, this can be attributed to the sudden suppression of 
Andreev reflection for electron energies beyond the sub-gap regime.

We see from the plots  Figs.~\ref{GvseV}(a)-(d) that the tunnelling conductance profile and the peak positions 
depend   significantly on the barrier strength. Nevertheless, the profile remains unchanged for barrier strengths which differ by 
 $\pi/2$, i.e., for $\chi\rightarrow\chi +\pi/2$.  This is understood by considering the simpler case of $\lambda=0$ for which the 
 reflection amplitude
at an arbitrary incident angle $\alpha_e$  has the expression
\begin{eqnarray}
r={\frac{(e^{2i\alpha{_e}}-1)\cos ({\beta}-2{\chi})}{2[\cos({\beta}-2{\chi})+i\cos({\alpha{_e}})\sin({\beta}-2{\chi})]}}.
\end{eqnarray}
The above expression remains invariant for every $\chi$ that differs by integer multiple of $\pi/2$. We also note that for  ${\lambda}=0$
 the peak value of TC is  $G/G_0=2$  [see Fig.~\ref{Two_figures}(a)]which  is achieved
when the reflection coefficient $r$ vanishes, or in other words  $|r_A|=1$ (follows from the unitarity criterion).  This is realised   for all incident angles
 when the transmission resonance criterion is satisfied, i.e., 
$\cos({\beta}-2{\chi})=0$.   In Fig.~\ref{Two_figures}(b)  we plot $eV/\Delta$ vs $\chi$ for which transmission resonance condition is satisfied.

The oscillatory behavior of $G/G{_0}$ as a function of $\chi$   persists even for non-zero $\lambda$'s. We plot this behavior
   in Figs.~\ref{Gvschi}(a)-(c), where the plots are for a fixed   $eV/\Delta$  and different $\lambda$'s.  As is expected, the peak value of $G/G_0=2$ is achieved for 
   incoming electrons whose energy is below the sub-gap regime.   However, the conductance can be made to switch off as illustrated in Fig.~\ref{Gvschi}(a) 
   with $\lambda= 0.8$, by tuning the out of-plane electric field to   $\lambda > eV/\Delta$. For $eV > \Delta$  absence of Andreev  reflection implies the
     peak value (for $\lambda=0$) to be at  most $G/G_0=1$  [see Fig.~\ref{Gvschi}(c)].    We note that   the oscillatory dependence of   $G/G{_0}$ on  $\chi$    is  in 
     complete contradiction to the normal metal-insulator-superconductor junctions, where increasing the barrier strength   always leads to the suppression of conductance~
     \cite{blonder1982transition}. 
  
\begin{figure}
\centering
\includegraphics[width=1.0\linewidth]{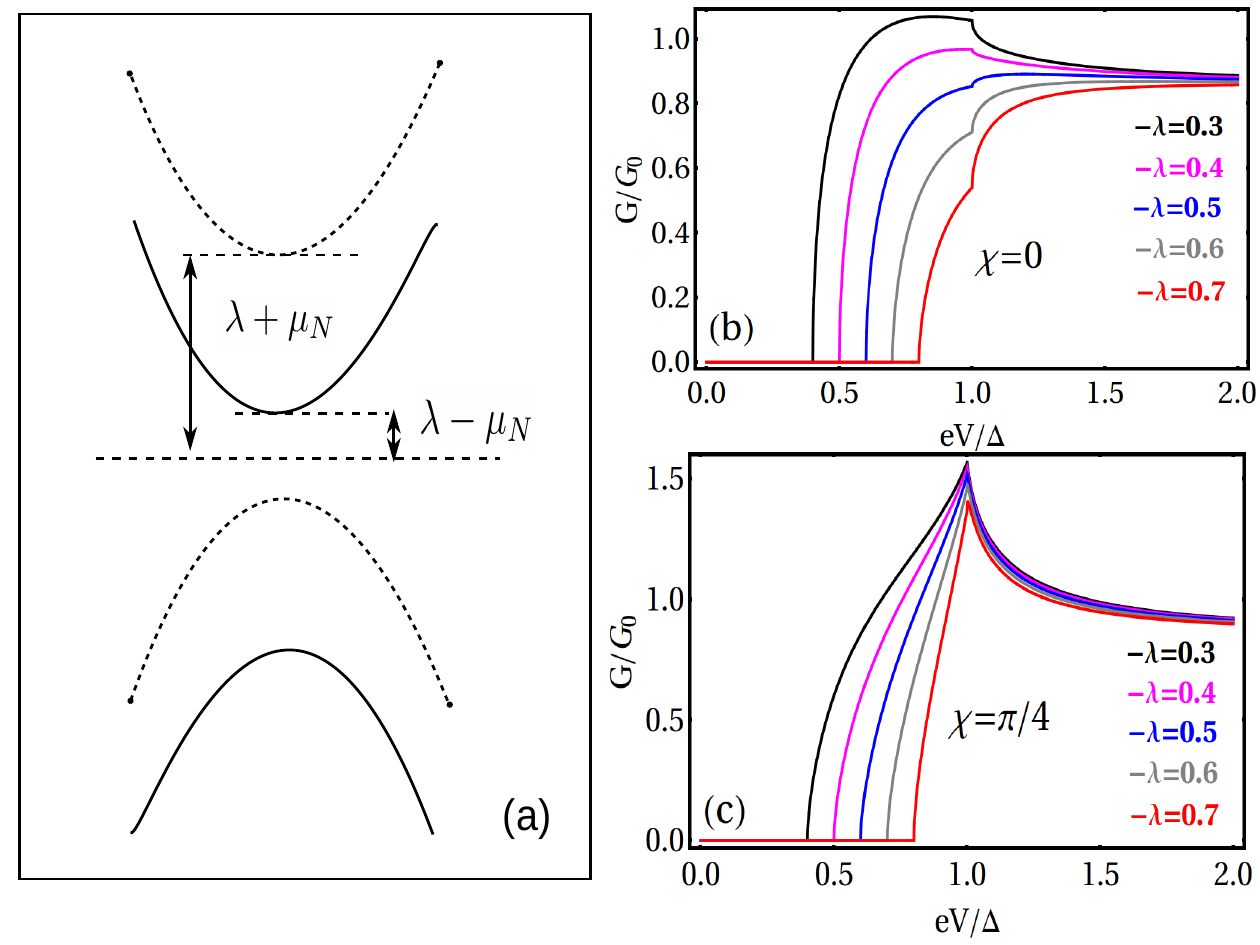}
\caption{(Color online)(a) BdG spectrum of  normal silicene  for $0<\mu_N/\Delta<\lambda<1$. Solid (dashed)  lines represent particle (hole) like spectrum.   
(b)-(c) The tunnelling conductance acquires non-zero values for 
$eV> \lambda\Delta + \mu_N$.}
\label{mu-lt-lambda}
\end{figure}

  For completeness we will consider  a scenario wherein the chemical potential is non-zero, 
  more specifically  $0<\mu_N/\Delta<\lambda<1$. 
    Now, although electronic levels  for $eV > \lambda \Delta  - \mu_N$ exist,
    yet the conductance remains zero until  the criterion  $eV > \lambda \Delta+ \mu_N$ is satisfied.
  This is due to the absence of Andreev reflection as there  are no states available for hole reflection [see Fig.~\ref{mu-lt-lambda}(a)]. We see this feature manifest itself in the $G/G_0$ vs $eV/\Delta$ plots 
 in Figs.~\ref{mu-lt-lambda}(b), (c).  Note 
 that unlike the $\mu_N=0$ case, now the transmission resonance criterion is not satisfied and the peak value of $G/G_0$ is smaller than $2$.
 
%
\begin{figure}[h]
\centering
\includegraphics[width=1.0\linewidth]{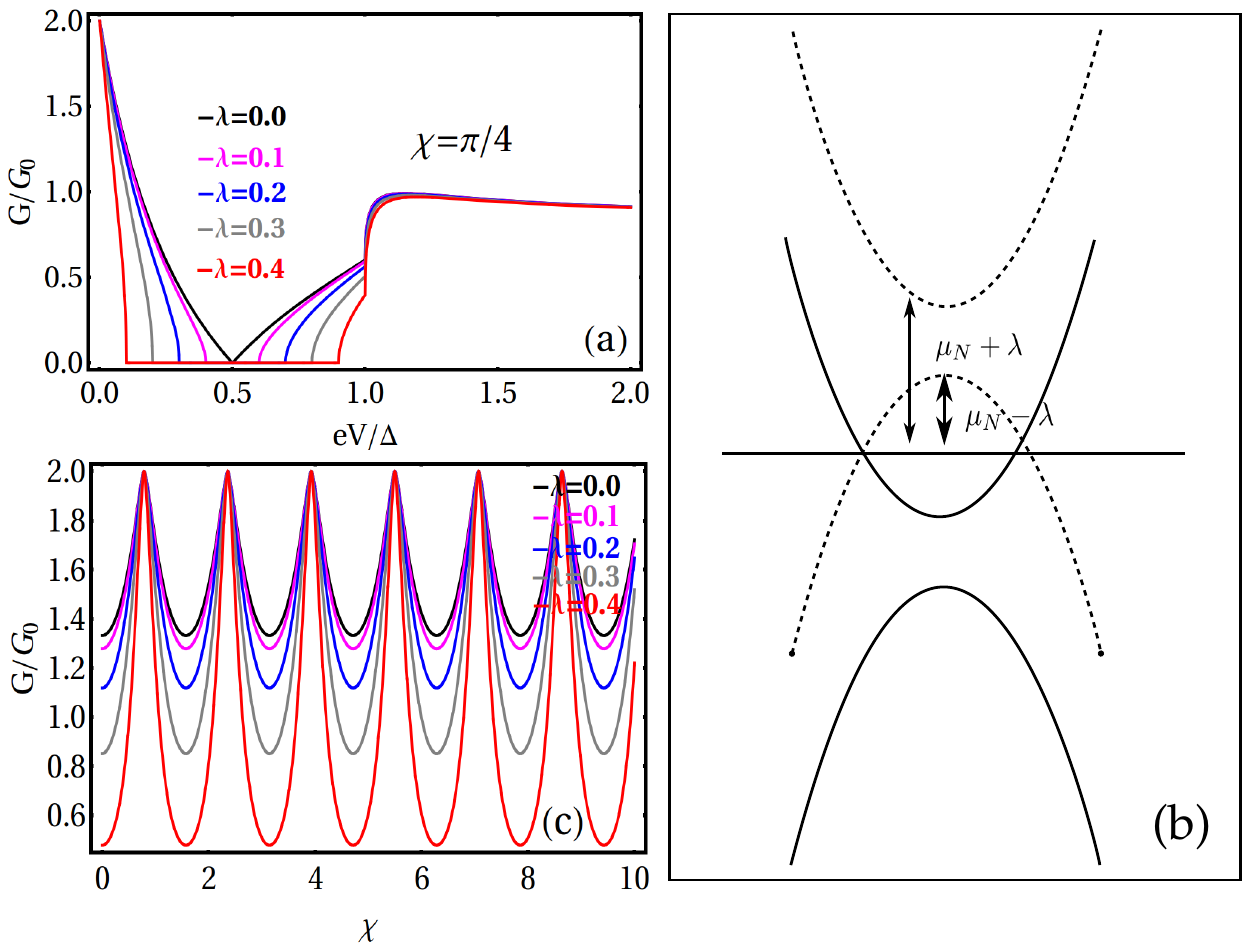}
\caption{(Color online)(a)  In the moderately doped regime $\mu_N =0.5\Delta$,  TC vanishes between the ranges $ \mu_N-\lambda\Delta<eV< \lambda\Delta+\mu_N$.   
(b) BdG spectrum for  $\lambda \Delta< \mu_N< l E_Z + \lambda_{SO}$. Andreev scattering is of retro type for $0< eV< \mu_N-\lambda\Delta$  and specular for $eV>\lambda\Delta+\mu_N $.
 (c)$G/G_0$ vs $\chi$  for $U_0 \gg \Delta$ exhibits  $\pi/2$ oscillations.}
\label{Fig6}
\end{figure}
%

\subsection{Moderately doped regime ($\mu_{N}\neq 0$)}
Here we will focus on moderate doping levels which we define as  $\lambda \Delta< \mu_N< l E_Z + \lambda_{SO}$ and set $\mu_N=0.5 \Delta$.   It turns out that the 
 profile of $G/G{_0}$ vs $eV$ plots  as shown in  Fig.~\ref{Fig6}(a)  are markedly  different from the earlier studied regimes.  We find  that the  conductance at  zero bias voltage starts out from a  non-zero 
value and monotonically decreases to zero at $eV=\mu_N-\lambda\Delta$. The conductance remains zero till $eV=\lambda\Delta+\mu_N$, beyond which it increases monotonically  till $eV=\Delta$.

For bias  voltages in the range $0< eV< \mu_N-\lambda\Delta$, the 
Andreev scattering is accompanied by hole scattering  of the usual retro type.  On the other hand, when an electron with energy $\mu_N-\lambda\Delta< eV<\lambda\Delta+\mu_N$ is incident on the interface of NS junction it gets completely reflected back  due to the absence of hole states [Fig.~\ref{Fig6}(b)].   
In a further twist,  an incident electron with energy in the range $eV>\lambda\Delta+\mu_N $ is again Andreev scattered due to the availability of hole states.
However, the holes now undergo specular reflection due to the change in sign of the curvature of hole spectrum. 

The  oscillatory feature in the  $G/G_0$ vs $\chi$ plots are shown in Fig.~\ref{Fig6}(c). For the simple case of $\lambda=0$  the 
expression for $r$ reduces to 
\begin{eqnarray}
r=\frac{i \left(-1+e^{2 i \alpha _e}\right) \sin ( 2 \chi-\beta )}{2 \cos (2 \chi - \beta ) \cos \left(\alpha
   _e\right)+2 i \sin (2 \chi -\beta)}.
\end{eqnarray}  
At zero bias, $\beta=\pi/2$,  the condition for transmission resonance reduces to ${\chi}=(n+1/2){\pi}/2$, so the first maxima 
is  exhibited at  ${\chi}={\pi}/4$.
\subsection{Highly doped regime}
In this subsection we present our tunneling conductance (TC)  results for the  highly doped regime i.e., $\mu_N\gg\Delta$.   The mean field criterion, $\mu_N +U_0 \gg \Delta$,
is now automatically satisfied irrespective of the value of $U_0$ . We will consider two scenarios, $U{_0}\gg\Delta$ and $U{_0}\ll\Delta$, and plot $G/G_0$ vs $eV/\Delta$ and vs $\chi$ in the two regimes.
Since $(lE_Z \pm \lambda_{SO})\sim \Delta$, both the bands in normal region are occupied and will contribute to the conductance. 

\begin{figure}
\centering
\includegraphics[width=1.0\linewidth]{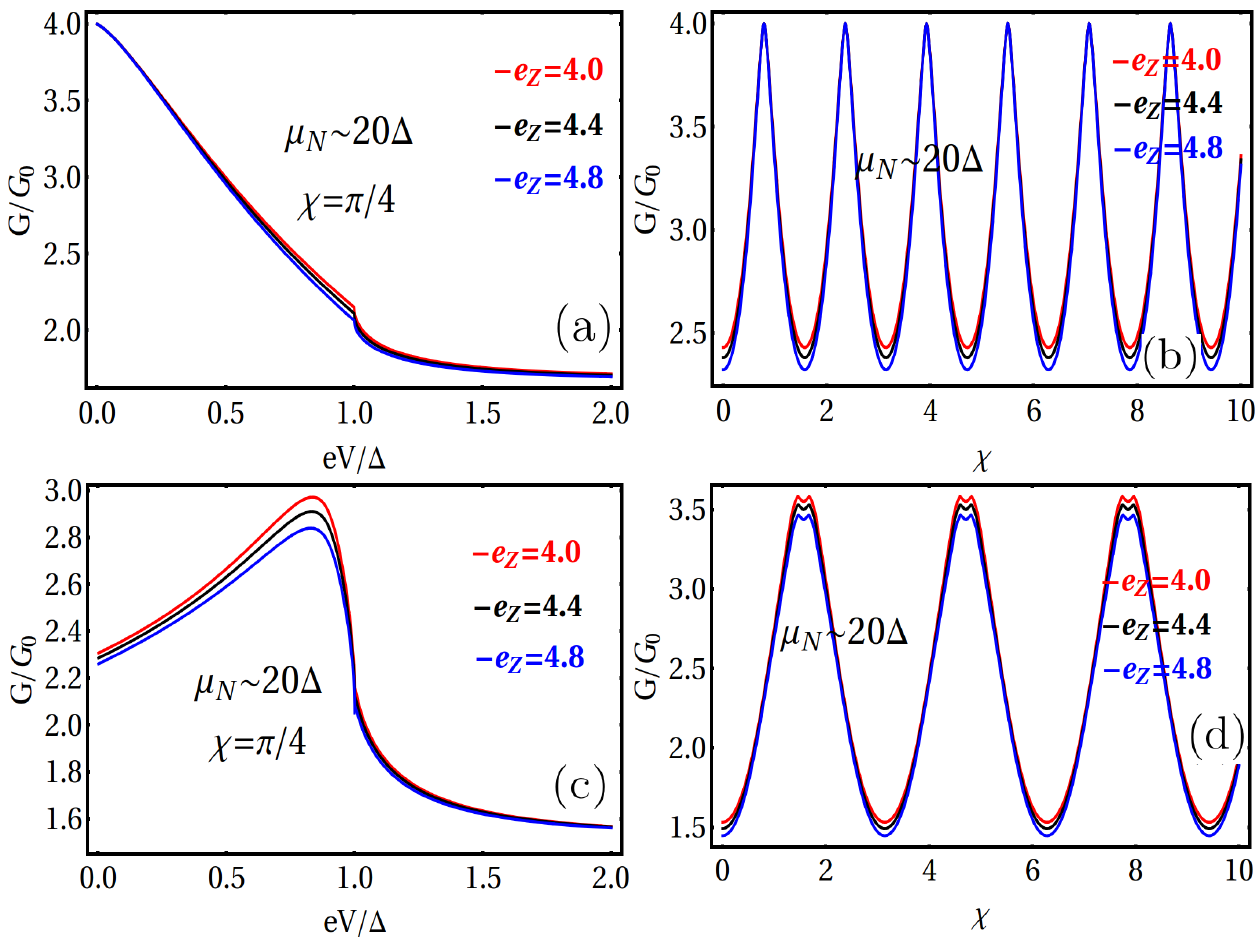}
\caption{(Color online)  Plot of tunneling conductance as a function of bias voltage [(a) and (c)], and barrier strength 
${\chi}$  [(b) and (d)]  with ${\lambda_{SO}}/{\Delta}=4.0$ and $e_Z= lE_Z/{\Delta}$. 
Figures (a) and (b) correspond to $U_0\gg \Delta$, (c) and (d)   correspond to $U_0\ll \Delta$. }
\label{Fig7}
\end{figure}

Due to the large value of the chemical potential the TC is nearly insensitive to the  variation in $\lambda$ and thus on the electric field 
applied perpendicular to the surface $E_Z$. However, the TC shows interesting behaviour
in the two extreme regimes for $U_0$. For large $U_0$ the TC exhibits, as before,  a change in the slope of
 $G/G_0$ vs $eV/\Delta$ curve at $eV=\Delta$ for all barrier strengths  [Fig.~\ref{Fig7}(a)]. Also,
 the  $\pi/2$ periodicity in the dependence of the TC  on the barrier strength persists  [Fig.~\ref{Fig7}(b)]. 
For small $U_0$ a similar change in slope at $eV=\Delta$ is present in   $G/G_0$ vs $eV/\Delta$ plots,  however, the TC now exhibits $\pi$ periodicity as a function of $\chi$  [see Figs.~\ref{Fig7}(c)-(d)].

When $U_0$ is large ($U{_0}\gg\Delta$), there is a large Fermi wave-length mismatch between the normal and the superconducting side. 
In this scenario, we obtain the $\pi/2$ periodicity in the dependence of the TC  on the barrier strength $\chi$. On the other hand, 
for small $U_0$ ($U{_0}\ll\Delta$), Fermi wave-length mismatch turns out to be negligible between the two sides. This gives rise to the $\pi$ periodicity~\cite{subhro}
in the behavior of TC with respect to $\chi$.

\section{Summary and conclusions} {\label{sec:IV}}
To summarize, in this article, we have presented a theory of tunneling conductance of a Normal-Insulator-Superconductor (NIS) 
junction of silicene in the thin barrier limit.  We have demonstrated that  in this limit the tunneling conductance shows novel periodic 
behavior as a function of barrier strength. In particular, we note that the period of oscillation changes from $\pi/2$ ($U_0\gg \Delta$) to $\pi$ ($U_0\ll \Delta$)
with the variation of doping in the superconducting side of silicene. Moreover, for the undoped regime ($\mu_{N}=0$), the external electric field
$E_{z}$ can be used as a switch to tune the conductance from on to off condition. The latter is a unique feature of silicene.

As far as experimental realization of our silicene NIS set-up is concerned, it can be possible to realize a proximity induced
superconducting gap in silicene by using $s$-wave superconductor like $\rm Al$~\cite{heersche2007bipolar}. Typical spin-orbit energy in silicene 
is $\lambda_{\rm SO}\sim 4~\rm meV$ while the buckling parameter $l\approx 0.23~\rm\AA$~\cite{CCLiu1,MEzawa3}. 
Considering Ref.~\onlinecite{heersche2007bipolar}, typical induced superconducting gap in silicene would be $\sim 0.2~\rm meV$. For such 
induced gap, the change of periodicity of TC from $\pi/2$ to $\pi$ may be possible to observe by changing the doping concentration 
from $\mu_{N}\sim 0.1~\rm meV$ to $\mu_{N}\sim 4~\rm meV$ for a barrier of thickness $\sim 10-15~\rm nm$ and height $V_{0}\sim 500~\rm meV$
which can be considered as thin barrier. Also the typical range of the external electric field can be within $E_{z}\sim 180-200~\rm V/\mu m$
to use our set-up as a switch.

We expect our results to be qualitatively similar to the recently discovered two-dimensional materials like
germenene, stanene~\cite{MEzawaReview,davila2014germanene,zhu2015epitaxial}. Although the strength of Rashba spin-orbit coupling
in these materials can be stronger than silicene~\cite{CCLiu1,MEzawa3}.

\vspace{2cm}

\appendix

\bibliography{Silicene_NIS_ref} 

\end{document}